\documentclass[12pt]{article}                   

\usepackage{graphicx}
\usepackage{hyperref}
\usepackage{amsmath}
\usepackage{amsthm}
\usepackage{amsfonts}
\usepackage{hyperref}
\usepackage[top=1in, bottom=1in, left=0.8in, right=0.8in]{geometry}
\usepackage[affil-it]{authblk}

\begin{document}

\title{Estimation theory and gravity}

\author{Can G\"{o}kler\footnote{Harvard University, Cambridge MA, USA} }

\date{\vspace{-7ex}}
\maketitle

\abstract
It is shown that if the Euclidean path integral measure of a minimally coupled free quantum scalar field on a classical metric background is interpreted as probability of 'observing' the field configuration given the background metric, then the maximum likelihood estimate of the metric satisfies Euclidean Einstein field equations with the stress-energy tensor of the 'observed' field as the source. In the case of a slowly varying metric, the maximum likelihood estimate is very close to its actual value. Then by virtue of the asymptotic normality of the maximum likelihood estimate, the fluctuations of the metric are Gaussian and governed by the Fisher information bi-tensor. Cramer-Rao bound can be interpreted as uncertainty relations between metric and stress-energy tensor. A plausible prior distribution for the metric fluctuations in a Bayesian framework is introduced. Using this distribution, the Euclidean decoherence functional acting on the field is calculated by integrating out the metric fluctuations around flat space.

\section{Introduction}
The main purpose of this paper is to investigate the mathematical relationship between estimation theory and Euclidean quantum field theory on curved background (Riemannian) space-time. We will see that, when the Euclidean path integral measure of a field propagating on a background metric is taken as the likelihood, the maximum likelihood of the metric satisfies the Euclidean Einstein equation with curvature square and cosmological constant terms. When the metric is slowly varying, the fluctuations are governed by the Fisher information bi-tensor, which can be directly expressed as a functional of the variance of the stress-energy tensor. Construction of a non-informative prior for the metric fluctuations leads to an action for fluctuations which could be directly expressed in terms of the Fisher information bi-tensor. When the fluctuations are integrated out, a non-local Euclidean decoherence functional is obtained for the field propagating on metric. 

The physical motivation of applying estimation theory to quantum fields on curved space-time stems from the following. One can postulate that the metric of space-time is not directly observable, but is inferred through its effects on the quantum fields that propagate on it. This yields uncertainties in the metric which are not due to the quantization of gravity, but the hidden nature of the metric. Even in the semi-classical quantum gravity, where the matter fields are treated as quantized but the gravitational field is treated classically, because of that the metric is not directly observable, there are probabilistic (non-quantum) uncertainties in the classical background metric. These uncertainties will in general depend on the measurements made on the matter field. In order to avoid the dependence on measurements, in this paper, we adopt the Euclidean path integral measure of a matter field propagating on a background metric as the likelihood. Therefore, the results of this paper do not directly correspond to the realistic Lorentzian setting with measurements on the matter field. However, the Euclidean setting provides us with an elegant mathematical framework where the connections between standard objects in estimation theory (e.g. maximum likelihood estimation, Fisher information, minimally information prior) and Euclidean quantum field theory on curved space-time can be made.

\section{Maximum likelihood estimator and gravitational field equations }

Consider the Euclidean action for the free scalar field $\phi$ on the background Riemannian metric $g_{\mu \nu}$ in local coordinates:

\begin{equation}
S_g[\phi]=\frac{1}{2}\int d^4x \sqrt{g}g_{\mu \nu} \partial^\mu \phi \partial^\nu \phi.
\end{equation}
For simplicity, we assume that the space-time manifold has compact topology and has no boundary. In this way we do not have to deal with the boundary conditions on $\phi$. Furthermore we assume that the manifold has positive Ricci curvature: $R > 3 \kappa >0$, which would yield an infrared cut-off scale. We note that, the constructions given below can in principle be generalized to settings where the manifold is non-compact, or has boundary and the field has mass, or is interacting, or has higher spin. Upon integration by parts, the action can be written as 

\begin{equation} \label{action}
S_g[\phi]=\frac{1}{2}\int d^4x \sqrt{g} \phi (-\Delta_g )\phi
\end{equation}
where $\Delta_g \cdot= 1 /\sqrt{g} \partial_\mu (\sqrt{g} g^{\mu\nu} \partial_\nu \cdot )$ is the Laplace-Beltrami operator associated with $g_{\mu \nu}$.  Since the manifold is compact and without boundary, $-\Delta_g$ is self-adjoint and have real non-negative discrete spectrum, i.e. for normalized eigenfunctions satisfying $-\Delta_g \theta_n= \lambda_n \theta_n$ and $\int  d^4x \sqrt{g} \theta_n \theta_m = \delta_{nm}$, $\lambda_0=0 < \lambda_1 \leq \lambda_2 \leq \cdots $ \cite{Schoen}. The lowest eigenvalue $\lambda_0$ corresponds to the constant function $\theta_0$. By Lichnerowicz theorem \cite{Schoen}, the next lowest eigenvalue $\lambda_1$ is bounded from below: $\lambda_1 > 4 \kappa$. This serves as the infrared cut-off scale.


We choose the conditional probability density of the field configuration $\phi$ given $g$ as the Euclidean path integral measure:

\begin{equation} \label{density}
p(\phi | g)=\frac{e^{-S_g[\phi]}}{Z_g}
\end{equation} 
where we set $\hbar=1$ and the partition function is $Z=\int \mathcal{D}\phi e^{-S_g[\phi]}$ with an appropriate measure $\mathcal{D}\phi$. Note that this is not a physical probability distribution. Nevertheless we follow the approach of \cite{Maloney} which used the above probability distribution to characterize proximities of quantum field theories. We think that a configuration of Euclidean field $\phi$ which could be 'observed'. We refer the reader to the last section on a discussion about choosing measurement independent probability distributions. 

We will show that, given the 'observed' field $\tilde{\phi}$, the maximum likelihood estimate of $g_{\mu \nu}$ satisfies Euclidean Einstein equations with quadratic curvature terms. In order to get finite answers, we need to invoke a renormalization scheme. We implement the scheme by fixing the volume and integrals of certain curvature polynomials over the space-time manifold to be finite. Specifically, we demand that 

\begin{equation} \label{volume}
\int d^4x \sqrt{g} = c_1,
\end{equation}

\begin{equation}
\int d^4x \sqrt{g} R = c_2,
\end{equation}

\begin{equation} \label{curvaturesquared}
\int d^4 x \sqrt{g} (R^{\mu\nu\rho\sigma}R_{\mu\nu\rho\sigma}-R^{\mu\nu}R_{\mu\nu}+30R^2+6\Delta_g R)=c_3
\end{equation}
where $c_1,c_2$ and $c_3$ are finite constants. Next, define the stress-energy tensor $T_{\mu \nu}$ as

\begin{equation}
T_{\mu \nu}(x)=-\frac{2}{\sqrt{g(x)}} \frac{\delta S_g[\phi]}{\delta g^{\mu\nu}(x)}.
\end{equation}
The effective action $W_g$ for $g_{\mu \nu}$ is defined by

\begin{equation}
e^{-W_g}=Z_g.
\end{equation}
Suppose a  particular Euclidean field configuration $\tilde{\phi}$ is 'observed'. The maximum likelihood estimate $\tilde{g}_{\mu \nu}$ of $g_{\mu \nu}$ is given by

\begin{equation}
\tilde{g}_{\mu \nu}= \text{argmax}_{g_{\mu \nu}} \log p(\tilde{\phi}| g)
\end{equation}
subject to the constraints determined by eq. \ref{volume}-\ref{curvaturesquared}. Noting that

\begin{equation}
\log p(\tilde{\phi}| g)=-S_g[\tilde{\phi}]+W_g
\end{equation}
and extremizing with respect to $g_{\mu \nu}$, with Lagrange multipliers $\delta_1, \delta_2$ and $\delta_3$ to incorporate the constraints, one obtains

\begin{align}
\frac{\delta W_g}{\delta g^{\mu\nu}(x)} \mid_{\tilde{g}} = & \frac{\delta S_g[\tilde{\phi]}}{\delta g^{\mu\nu}(x)} \mid_{\tilde{g}} + \frac{\delta}{\delta g^{\mu\nu}(x)} \mid_{\tilde{g}} [ \delta_1 (\int d^4x \sqrt{g} - c_1)  + \delta_2 (\int d^4x \sqrt{g} R - c_2) \\ \nonumber &  +\delta_3  (\int d^4 x \sqrt{g} (R^{\mu\nu\rho\sigma}R_{\mu\nu\rho\sigma}-R^{\mu\nu}R_{\mu\nu}+30R^2+6\Delta_g R) - c_3) ].
\end{align}
The left hand side is given more explicitly as

\begin{equation} \label{effective-stress}
\frac{\delta W_g}{\delta g^{\mu\nu}(x)} =-\frac{1}{Z_g} \frac{\delta Z_g}{\delta g^{\mu \nu}(x)}= \int \mathcal{D} \phi \frac{\delta S_g[\phi]}{\delta g^{\mu\nu}(x)} \frac{e^{-S_g[\phi]}}{Z_g} = \langle \frac{\delta S_g[\phi]}{\delta g^{\mu\nu}(x)} \rangle = - \frac{\sqrt{g(x)}}{2} \langle T_{\mu \nu}(x)\rangle
\end{equation}
where $\langle f \rangle$ denotes the expectation of $f$ over $p(\phi |g)$. Hence

\begin{align} \label{einstein}
T_{\mu\nu}\mid_{\tilde{g}, \tilde{\phi}}  = & \langle T_{\mu\nu}\rangle \mid_{\tilde{g}} + \frac{2}{\sqrt{g(x)}} \frac{\delta}{\delta g^{\mu\nu}(x)} \mid_{\tilde{g}} [ \delta_1 \int d^4x \sqrt{g}   + \delta_2 \int d^4x \sqrt{g} R \\ \nonumber &  +\delta_3  \int d^4 x \sqrt{g} (R^{\mu\nu\rho\sigma}R_{\mu\nu\rho\sigma}-R^{\mu\nu}R_{\mu\nu}+30R^2+6\Delta_g R) ].
\end{align}
On the left hand side of this equation, we have the stress-energy tensor of the observed field on the background $\tilde{g}$. The right hand side will yield the Einstein tensor together with the cosmological constant and a term coming from terms quadratic in curvature in the action after choosing $\delta_1, \delta_2$ and $\delta_3$ to cancel the divergences in $\langle T_{\mu\nu}\rangle \mid_{\tilde{g}}$. In this sense, eq. $\ref{einstein}$ is the Euclidean Einstein equation. 

Next, we calculate $\langle T_{\mu\nu}\rangle \mid_{\tilde{g}}$. This is a a standard problem in quantum field theory in curved space-time \cite{Hawking, Davies, Wald1979}. Since $p(\phi|g)$ is Gaussian, the partition function $Z_g$, therefore the effective action $W_g$, can be expressed in terms of $\Delta_g$ as follows. Put the eigenfunction expansion $\phi=\sum_{n=1}^\infty c_n \theta_n$ in the action (eq. \ref{action}) to obtain

\begin{equation}
S_g[\phi]=\frac{1}{2} \sum_{n=1}^\infty \lambda_n c_n^2.
\end{equation}
Note that we omitted the $\theta_0$ term since it does not contribute to the action. Choose the measure $\mathcal{D}\phi= \prod_{n=1}^\infty \frac{dc_n}{\sqrt{2 \pi}}$. Then, $Z_g$ is a Gaussian integral. One computes $W_g$ as

\begin{equation}
W_g = -\log \int \mathcal{D}\phi  e^{-S_g[\phi]} = -\log \prod_{n=1}^\infty \frac{dc_n}{\sqrt{2 \pi}} e^{-\frac{1}{2} \sum_{n=1}^\infty \lambda_n c_n^2} = \frac{1}{2} \log  \prod_{n=1}^\infty \lambda_n= \frac{1}{2} \log \det (-\Delta_g)
\end{equation}
where we defined $\det (-\Delta_g)= \prod_{n=1}^\infty \lambda_n$. Using Schwinger time formalism \cite{Visser}, we can further express $W_g$ as

\begin{equation} \label{effectivefinal}
W_g=\frac{1}{2} \log \det (-\Delta_g)= \frac{1}{2} \text{tr} \log (-\Delta_g) = - \frac{1}{2} \int_0^\infty \frac{ds}{s} \text{tr} ( e^{s \Delta_g} )+c
\end{equation}
where $c$ is an infinite constant and $\text{tr}(\cdot)$ denotes the operator trace. Note that 

\begin{equation}
\text{tr} ( e^{s \Delta_g}) = \int d^4 x \sqrt{g} \sum_{n=1}^\infty \theta_n(x) e^{s\Delta_g} \theta_n(x)=\sum_{n=1}^\infty e^{-\lambda_n s}.
\end{equation}
Since $\lambda_n \geq \lambda_1 > 4 \kappa$ for $n \geq 1$, in the $s$ integral in eq. \ref{effectivefinal}, the terms with $s \gg 1/(4 \kappa)$ are negligible. Therefore

\begin{equation}
W_g = - \frac{1}{2} \int_0^{1/ \kappa} \frac{ds}{s} \text{tr} ( e^{s \Delta_g} )+c.
\end{equation}
To regularize the integral, we also introduce a lower cut-off at $s=1 / k^2$:

\begin{equation} \label{effective2}
W_g = - \frac{1}{2} \int_{1/ k^2}^{1/\kappa} \frac{ds}{s} \text{tr} ( e^{s \Delta_g} )+c.
\end{equation}
The contributions to the integral from the upper limit is finite and negligible as compared to the diverging contributions arising from the lower limit \cite{Hawking, Davies, Wald1979}. To calculate the integral, we will expand the heat kernel $e^{s \Delta_g}$ for small $s$. Since the space-time is compact,  one can expand $e^{s \Delta_g}$ as

\begin{equation}
\text{tr}(e^{s \Delta_g})= \int d^4 x\frac{\sqrt{g}}{(4 \pi s)^2} (b_0(g)+b_1(g)s+b_2(g)s^2+O(s^3))
\end{equation}
where the coefficients are expressed in terms of the curvature tensor: $b_0=1$, $b_1=\frac{R}{6}$ and $b_2=\frac{1}{720}(R^{\mu\nu\rho\sigma}R_{\mu\nu\rho\sigma}-R^{\mu\nu}R_{\mu\nu}+30R^2+6\Delta_g R)$\cite{Hawking, Davies, Wald1979}. Using this expansion in eq. \ref{effective2}, we obtain

\begin{align}
W_g =- \frac{1}{32 \pi^2} \int d^4 x \sqrt{g} (\frac{k^4}{2}  +\frac{k^2}{6} R +\log(k^2)b_2)+O(\frac{1}{k^2})+c.
\end{align}
Substituting this into eq. \ref{einstein} and using eq. \ref{effective-stress}, we get

\begin{align} \label{einstein2}
T_{\mu\nu}\mid_{\tilde{g}, \tilde{\phi}}  = & \frac{2}{\sqrt{g(x)}} \frac{\delta}{\delta g^{\mu\nu}(x)} \mid_{\tilde{g}} [ \frac{1}{32 \pi^2} \int d^4 x \sqrt{g} (\frac{k^4}{2}  +\frac{k^2}{6} R +\log(k^2)b_2) \\ \nonumber & +  \delta_1 \int d^4x \sqrt{g}   + \delta_2 \int d^4x \sqrt{g} R   +\delta_3  \int d^4 x \sqrt{g} 720 b_2 ].
\end{align}
Next, we renormalize the right hand size by choosing $\delta_1(k), \delta_2(k)$ and $\delta_3(k)$ to cancel the $k$ dependent terms which diverge as $k \rightarrow \infty$. After the divergences are canceled, eq. \ref{einstein2} assumes the form

\begin{equation} \label{einstein3}
T_{\mu\nu}\mid_{\tilde{g}, \tilde{\phi}}  =  \frac{2}{\sqrt{g(x)}} \frac{\delta}{\delta g^{\mu\nu}(x)} \mid_{\tilde{g}}  \int d^4 x \sqrt{g} ( \alpha_1 + \alpha_2 R + \alpha_3 (R^{\mu\nu\rho\sigma}R_{\mu\nu\rho\sigma}-R^{\mu\nu}R_{\mu\nu}+30R^2+6\Delta_g R) )
\end{equation}
On the left hand side, we have the observed stress energy tensor. On the right hand side, we have the Einstein tensor together with the cosmological constant and curvature squared terms. Therefore Euclidean Einstein-like equation arises from maximum likelihood estimation. After choosing $\alpha_1,\alpha_2$ and $\alpha_3$, given the observed field $\tilde{\phi}$, one can solve eq. \ref{einstein3} for $\tilde{g}_{\mu \nu}$. Using this $\tilde{g}_{\mu \nu}$, one can calculate the constants $c_1, c_2$ and $c_3$ defined by eqs. \ref{volume}-\ref{curvaturesquared}. We remark that any choice of $\alpha_1, \alpha_2$ and $\alpha_3$ is compatible with the maximum likelihood procedure with a suitable choice of constants $c_1, c_2$ and $c_3$, as long as eq. \ref{einstein3} has a solution.

\section{Fluctuations and Fisher information}

From now on, we assume that the radius of curvature is large as compared to typical wavelength of the matter field, or the metric is slowly varying compared to the matter field. To ensure this, one can assume that the Ricci curvature is bounded from above. Therefore, when one observes the matter field, several observation points give knowledge about the same representative point for the metric field. In this way, we can think that $p(\phi | g)$ represents a large sample likelihood. As known from asymptotic statistics, the maximum likelihood estimator is consistent \cite{Vaart}, which means that as the number of data points get large, the maximum likelihood estimator converges to the true value of the parameter (in our case the metric). Thus, as there are more points in the matter field for each point in the metric (one can think of a lattice or simplicial discretization so that one lattice point of the metric corresponds to many lattice points of matter, or the metric is effectively constant throughout a large number of points), the result of maximum likelihood inference should match with the actual value of the metric. In this sense the Euclidean Einstein equation (eq. \ref{einstein3}) governs the evolution of the true underlying metric $g_0$, as the maximum likelihood estimated $\tilde{g}$ converges to $g_0$ in the large sample limit. While $\tilde{g}$ converges to $g_0$, the fluctuations $\Delta g$ around $g_0$ are governed by the Fisher information tensor \cite{Vaart}. The Fisher information at $g_0$ is the 4 index object (bi-tensor) on the two copies of the space-time manifold

\begin{equation}
F_{\mu\nu\rho\sigma}(g_0)(x,y)= \int \mathcal{D} \phi p(\phi | g) \frac{\delta \log p(\phi | g)}{\delta g^{\mu\nu} (x)}\frac{\delta\log p(\phi | g)}{\delta g^{\rho\sigma}(y)} |_{g_0}.
\end{equation}
$F$ can be expressed in terms of the stress-energy tensor as follows. Compute

\begin{equation}
 \frac{\delta \log p(\phi | g)}{\delta g^{\mu\nu} (x)}=-\frac{\delta S_g[\phi]}{\delta g^{\mu\nu}(x)}- \frac{1}{Z_g} \frac{\delta Z_g}{\delta g^{\mu \nu}(x)}
 \end{equation} 
where the expectation value $\langle \cdot \rangle$ is taken with respect to $p(\phi | g)$. Hence

\begin{equation}
F_{\mu\nu\rho\sigma}(g_0)(x,y)= \frac{1}{4} \sqrt{g_0(x)} \sqrt{g_0(y)} \langle (T_{\mu\nu}(x)- \langle T_{\mu\nu}(x) \rangle )(T_{\rho\sigma}(y)- \langle T_{\rho\sigma}(y) \rangle) \rangle |_{g_0}.
\end{equation}
One can use the Fisher information bi-tensor to get lower bounds on fluctuations of the metric. As a motivation, consider the energy-time uncertainty relation in non-relativistic quantum mechanics:

\begin{equation}
\Delta t^2 \Delta E^2 \geq \frac{\pi^2 \hbar^2}{4}
\end{equation}
where $\Delta E^2=\sqrt{\langle H^2 \rangle-\langle H \rangle^2}$ is the variance of energy when the dynamics is generated by the Hamiltonian $H$. In a space-time with metric $g_{\mu\nu}$, $\Delta t = \Delta g_{00}$
and $\Delta E= \Delta T_{00}$. Vectorize $\mu \nu$ and $\rho \sigma$ indices as $i$ and $j$, respectively in $F_{\mu\nu\rho\sigma}(g_0)(x,y)=F_{ij}(g_0)(x,y)$. Cramer-Rao bound\cite{Cover} (for an unbiased estimator) provides a relativistic generalization of the uncertainty relation: 

\begin{equation} \label{Cramer}
\langle \Delta g_i(x) \Delta g_j(y) \rangle \geq [F_{ij}(g_0)(x,y)]^{-1}=\frac{4}{\sqrt{g_0(x)} \sqrt{g_0(y)}} [\langle \Delta T_i(x) \Delta T_j(y) \rangle |_{g_0}]^{-1}
\end{equation}
where $[ \cdot ]^{-1}$ denotes the matrix inverse, $\Delta g_{\mu\nu}(x)=g_{\mu\nu}(x)-\langle g_{\mu\nu}(x)\rangle=g_{\mu\nu}(x)-g_0(x)$ and $\Delta T_{\mu\nu}(x)=T_{\mu\nu}(x)-\langle T_{\mu\nu}(x)\rangle$. Here, the estimator used in the Cramer-Rao bound is the maximum likelihood estimator introduced in the last section. The maximum likelihood estimator is asymptotically unbiased \cite{Vaart}, i.e. the bias goes to zero as the number of samples grow. And as discussed before, it converges to the true metric $g_0$. Therefore, in eq. \ref{Cramer}, we used $g_0$ in the place of $\tilde{g}$. The Cramer-Rao bound can be used to derive a weaker bound\cite{Bobrovsky} for the components of $\Delta g$. For instance

\begin{align}
\langle \Delta g_{00}(x) \Delta g_{00}(y) \rangle \geq ([F_{ij}(g_0)(x,y)]^{-1})_{00} \geq (F_{ij}(g_0)(x,y))_{00})^{-1} \nonumber
\\ =\frac{4}{\sqrt{g_0(x)} \sqrt{g_0(y)}} [\langle \Delta T_{00}(x) \Delta T_{00}(y) \rangle |_{g_0}]^{-1}.
\end{align}
We remark that similar uncertainty relations were obtained by \cite{Downes}. The above equation looks very much like the standard energy-time uncertainty relation. We elaborate on the statistical interpretation of this uncertainty relation as follows assuming that standard concepts of multi-dimensional estimation theory applies to our case. Indeed, when we discretize the metric and the field, we are in very high but still finite dimensional setting and the results from estimation theory should apply. We think $\tilde{g}_{\mu\nu}(x)(\phi)$ as an estimator, say the maximum likelihood estimator: a function of the random variable $\phi$, the field configuration.  If the estimator is efficient (satisfies the Cramer-Rao bound), the uncertainty relation is satisfied by equality. If the Fisher information is large (it diverges indeed), then the estimator has to be only asymptotically efficient (efficient as the number of samples go large). The inverse covariance of metric fluctuations around the true metric (this is the background metric) is given by Fisher information where the estimator is assumed to be consistent (converges to the true value with large Fisher information). There is a fixed background metric which is to be estimated. However, it cannot be 'observed' directly. The information about the fluctuations of geometry comes from matter fields. Fisher information is the fundamental limitation to the accuracy to which any observer can resolve the metric. Therefore metric is fluctuating at the rate determined by the Fisher information in the large sample limit. 

\section{Minimally informative prior for the metric}

Given the conditional density $p(\phi | g)$, suppose we would like to know how we can construct a prior distribution for the background metric $g$. The background metric cannot be 'observed' directly. Therefore, in order to make predictions, one must marginalize $p(\phi, g)$ over $g$ to get $p(\phi)$: if we regard $g$ as the background field which cannot be observed directly, all the observable predictions of the theory is determined by $p(\phi)$. The problem is then to find an objective prior $p(g)$, given only the conditional density (likelihood) $p(\phi | g)$. Above, the action for fluctuations is shown to be determined by Fisher information. Then for metric fluctuations, $p(g)$ is Guassian, centered on the estimate and with inverse covariance matrix of fluctuations given by the Fisher information. In the asymptotic limit (large Fisher information), the Bayesian procedure converges to a normal distribution with the above properties. This is the Laplace-Bernstein-Von Mises-Le Cam theorem on asymptotic normality\cite{Vaart}. This holds for finite dimensional problems with mild assumptions on the prior (it should have non-zero probability around a neighbourhood of the true parameter). For infinite dimensional (non-parametric) problems which one faces in the case of field theory, this issue should be more delicate. However, if we descretize the fields, then the finite dimensional results should apply. From now on, we assume that finite dimensional asymptotic normality results apply to the space of metrics. A variational method exists (which goes with the name 'reference prior' in literature\cite{Bernardo, Berger}), which yields a prior $p(g)$ in the form that is sought by the form of the gravitational action. The variational principle is to maximize the mutual information between $g$ and $\phi$:

\begin{equation}
p(g)= \text{argmax}_{p(g)} I(g, \phi)
\end{equation}
where the mutual information is defined by

\begin{equation}
I(g, \phi)=\int \mathcal{D}g\mathcal{D}\phi p(g, \phi) \log\frac{p(g, \phi)}{p(g) p(\phi)}.
\end{equation}
We can interpret such a choice of prior as that the gravitational field reacts to the matter field to maximize the information revealed in the matter field about it. The dynamics of gravity is determined by maximizing the correlations between matter fields and the gravitational field. The matter fields enable an observer to get as much information as possible about the background metric. Note that this choice of prior is a realization of principle of indifference: one chooses the prior of least information if at the end one acquires maximum information. The mutual information is a concave function of p(g) (at least in the finite dimensional setting), therefore one expects a unique maximum. However, it is in general hard to compute the maximum. There is one exception. If the posterior distribution $p(g | \phi)$ is independent of $\phi$, then the maximization is straightforward. But, one knows from asymptotic normality that this holds. So if the Fisher information is sufficiently large, $p(g | \phi)$ is independent of $\phi$ and one can compute $p(g)$ with relative ease. To see this, write $I(g, \phi)$ in the following form

\begin{equation}
I(g, \phi)=\int \mathcal{D}g p(g) \log \frac{e^{\int \mathcal{D} \phi p(\phi | g) \log p(g|\phi)}}{p(g)}.
\end{equation}
Define $f(g)=e^{\int \mathcal{D} \phi p(\phi | g) \log p(g|\phi)}$. If $p(g | \phi)$ is independent of $\phi$, then $f(g)$ does not depend on $p(g)$. In this case, the extremum of $I(g, \phi)$ occurs when $p(g) \propto f(g)$. To calculate $f(g)$ one needs the posterior. Asymptotic normality tells that

\begin{equation}
p(g | \phi) \propto e^{-\frac{1}{2} \int d^4 x \int d^4 y \int F_{\mu\nu\rho\sigma}(\hat{g})(x,y) (g^{\mu\nu}(x)-\hat{g}^{\mu\nu}(x)(\phi)) (g^{\rho\sigma}(y)-\hat{g}^{\rho\sigma}(y)(\phi)) }
\end{equation}
where $g_0$ is the true background metric and $\hat{g}_{\mu\nu}(x)(\phi)$ is a consistent estimator of $g_{\mu\nu}$ given the observed configuration $\phi$ (converges to $g_0$ in probability $p(\phi | g)$). Using this, it follows that

\begin{equation}
p(g) \propto e^{-\frac{1}{2} \int d^4 x \int d^4 y \int F_{\mu\nu\rho\sigma}(\hat{g})(x,y) \langle (g^{\mu\nu}(x)-\hat{g}^{\mu\nu}(x)(\phi)) (g^{\rho\sigma}(y)-\hat{g}^{\rho\sigma}(y)(\phi))\rangle_{p(\phi | g)}}
\end{equation}
where $\langle \cdot \rangle_{p(\phi | g)} $ denotes the expectation taken with respect to $p(\phi | g)$. In the asymptotic limit one can let $\hat{g}_{\mu\nu}(x)(\phi)=(g_0)_{\mu \nu}$.  Since $\hat{g}^{\mu \nu}(\phi) \hat{g}_{\nu \sigma}(\phi)=\delta^\mu_\sigma$ must be satisfied for all $\phi$, the inverse $\hat{g}^{\mu \nu}(\phi)$ of the estimator $\hat{g}_{\nu \sigma}(\phi)$ must converge to the inverse $g_0^{\mu \nu}$ of the true metric $(g_0)_{\mu \nu}$. Hence, the above $p(g)$ has the form $p(g) \propto e^{-S[g]}$, with $S[g]$ the gravitational action for the fluctuations:

\begin{equation} \label{fluctuations}
p(g) \propto e^{-\frac{1}{2} \int d^4 x \int d^4 y \int F_{\mu\nu\rho\sigma}(g_0)(x,y) (g^{\mu\nu}(x)-g_0^{\mu\nu}(x)) (g^{\rho\sigma}(y)-g_0^{\rho\sigma}(y))}.
\end{equation}

\section{Euclidean decoherence functional in flat space}

We would like integrate out the metric fluctuations to calculate the Euclidean decoherence functional acting on the field. To do this consider fluctuations $h_{\mu\nu}$ around the Euclidean space with metric $\delta_{\mu\nu}$. The total metric has the form $g_{\mu\nu}=\delta_{\mu\nu}+h_{\mu\nu}$. Expanding $S_g[\phi]$ to first order in $h_{\mu\nu}$ we obtain

\begin{equation}
S_g[\phi]=\frac{1}{2}\int d^4x(\partial^\mu \phi \partial_\mu \phi)+\frac{1}{2} \int d^4x h_{\mu\nu}(\partial^\mu \phi \partial^\nu\phi+\frac{1}{2}\delta^{\mu\nu}\partial^\sigma \phi \partial_\sigma \phi)+O(h^2).
\end{equation}
To get the effective action for the field, we formally (not paying attention to gauge redundancies in $h_{\mu\nu}$) integrate out the Gaussian metric fluctuations using eq. \ref{fluctuations}:
\begin{align}
e^{-S_{\text{eff}}[\phi]} & = \int \mathcal{D}g p(g) \frac{e^{-S_g[\phi]}}{Z_g} \nonumber
\\ & \propto e^{-\frac{1}{2}\int d^4x(\partial^\mu \phi \partial_\mu \phi )+\frac{1}{4}\int d^4x \int d^4y F^{-1}_{\mu\nu\rho\sigma}(x,y)A^{\mu\nu}(\phi(x))A^{\rho\sigma}(\phi(y))}
\end{align}
where $A^{\mu\nu}(\phi(x))=(\partial^\mu \phi \partial^\nu\phi+\frac{1}{2}\delta^{\mu\nu}\partial^\sigma \phi \partial_\sigma \phi)$. Note that we omitted the contributions from $Z_g$ supposing that the fluctuations in the metric dominate over the quadratic terms that appear in the expansion of $Z_g$ in powers of $h_{\mu\nu}$. We see that $S_{\text{eff}}[\phi]$ can be written as the sum of the action of the scalar field in Euclidean space $S_0[\phi]$ and a non-local decoherence term $S_d[\phi]$ quartic in fields:

\begin{equation}
S_{\text{eff}}[\phi]=S_0[\phi]+S_d[\phi]
\end{equation}
If the field is conformally coupled, which would modify $A^{\mu \nu}$, one can obtain an explicit expression for the Fisher information bi-tensor using its relation to the stress energy tensor as given in eq. 21 \cite{Maloney, Osborn}:

\begin{equation}
F_{\mu\nu\rho\sigma}(x,y) = C \frac{I_{\mu\nu\rho\sigma}(x-y)}{|x-y|^8}
\end{equation}
where $I_{\mu\nu\rho\sigma}(x)=\frac{1}{2} (I_{\mu\sigma}(x)I_{\nu\rho}(x)+I_{\mu\rho}(x)I_{\nu\sigma}(x))-\frac{1}{4}\delta_{\mu\nu}\delta_{\sigma\rho}$, $I_{\mu\nu}(x)=\delta_{\mu\nu}-2\frac{x_\mu x_\nu}{|x|^2}$ and $C$ is a constant. As in section 2, one can vectorize and invert $F_{\mu\nu\rho\sigma}$ to get an explicit expression for the decoherence functional but we do not pursue this calculation here.

The Euclidean decoherence functional obtained above is an inevitable consequence of our inability to directly observe the gravitational field. When one takes into account the fluctuations in the metric due to our lack of knowledge, averaging over the metric fluctuations results in a decoherence term for the matter field. The decoherence due to gravitational fluctuations is not a new idea and have been explored in the context of spontaneous collapse models \cite{Diosi, Penrose, Bassi}. In principle, in this paper, we have a Euclidean relativistic version of Di\'{o}si's original argument that the origin of gravitational fluctuations which induce collapse is the limitations to the measurability of the metric by quantum probes. 

\section{Scholia}

\indent

In this paper, we used the unphysical Euclidean measure as the conditional probability distribution. We used it to avoid the dependence of the probabilities on specific measurements. However, there are other objective probability distributions, which do not depend on particular measurements. For example ,in principle one can consider a continuous measurement of the matter field and maximize a measure of information about the metric such as the mutual information between the metric and matter field over all continuous measurements. Incorporations of continuous measurements into path integral formalism can be found for instance in \cite{Caves, Mensky}. Another way is to start from a specific state of the matter field and make a measurement at a certain predetermined time and maximize the classical Fisher information over all such measurements, therefore obtaining the quantum Fisher information as the measure of fluctuations of the metric. A more unconventional objective probability distribution can be constructed via Nelson's stochastic formulation of quantum mechanics \cite{Nelson, Guerra}. In this formulation, to each wave function evolution, one associates a Markovian stochastic process in the configuration space of the matter fields. The path measure of this stochastic process can serve as the conditional probability. 


Derivations of Einstein equations from results in quantum field theory and statistical principles are well known \cite{Jacobson, Verlinde, Lloyd}. For example, Jacobson \cite{Jacobson} showed that assuming area law for entropy, Unruh effect and the thermodynamic equation of state, one can derive the semi-classical Lorentzian Einstein equations. To compare, we assume quantum field theory on curved spacetime which would imply the area law and the Unruh effect, and instead of the thermodynamic equation of state, we have the principle of maximum likelihood estimation.

\section*{Acknowledgements}
We thank Vijay Balasubramanian, Can Koz\c{c}az, Nima Lashkari, Seth Lloyd and Alexander Maloney for suggestions and discussions. We thank the anonymous referee for suggestions.

\bibliographystyle{unsrt}
\bibliography{estimation_gravity_references}

\begin{thebibliography}{10}

\bibitem{Schoen}
R.~Schoen and S.T. Yau.
\newblock {\em Lectures on Differential Geometry}.
\newblock International Press of Boston, 1994.

\bibitem{Maloney}
V.~Balasubramanian, J.J. Heckman, and A.~Maloney.
\newblock Relative entropy and proximity of quantum field theories.
\newblock {\em Journal of High Energy Physics}, 2015:104, Jul 2015.

\bibitem{Hawking}
S.~Hawking.
\newblock Zeta function regularization of path integrals in curved spacetime.
\newblock {\em Communications in Mathematical Physics}, 55:133--148, 1977.

\bibitem{Davies}
N.D. Birrell and P.~C.~W. Davies.
\newblock {\em Quantum fields in curved spacetime}.
\newblock Cambridge University Press, 1984.

\bibitem{Wald1979}
Robert~M Wald.
\newblock {On the Euclidean approach to quantum field theory in curved
  spacetime}.
\newblock {\em Communications in Mathematical Physics}, 70(3):221--242, 1979.

\bibitem{Visser}
M.~Visser.
\newblock Sakharov's induced gravity: a modern perspective.
\newblock {\em Modern Physics Letters A}, 17:977--991, 2002.

\bibitem{Vaart}
A.~W. van~der Vaart.
\newblock {\em Asymptotic Statistics}.
\newblock Cambridge Series in Statistical and Probabilistic Mathematics.
  Cambridge University Press, 1998.

\bibitem{Cover}
T.M. Cover and J.A. Thomas.
\newblock {\em Element of information theory, 2nd Ed.}
\newblock Wiley, 2006.

\bibitem{Bobrovsky}
B.~Z. Bobrovsky, E.~Mayer-Wolf, and M.~Zakai.
\newblock Some classes of global cramer-rao bounds.
\newblock {\em The Annals of Statistics}, 15(4):1421--1438, 1987.

\bibitem{Downes}
T.~G. Downes, G.~J. Milburn, and C.~M. Caves.
\newblock Optimal quantum estimation for gravitation.
\newblock page arXiv:1108.5220, 2011.

\bibitem{Bernardo}
J.M. Bernardo.
\newblock Reference posterior distributions for bayesian inference.
\newblock {\em Journal of the Royal Statistical Society: Series B
  (Methodological)}, 41:113--128, Jan 1979.

\bibitem{Berger}
James~O. Berger, José~M. Bernardo, and Dongchu Sun.
\newblock The formal definition of reference priors.
\newblock {\em The Annals of Statistics}, 37(2):905--938, 2009.

\bibitem{Osborn}
H.~Osborn and A.~Petkou.
\newblock Implications of conformal invariance in field theories for general
  dimensions.
\newblock {\em Annals of Physics}, 231(2):311–362, May 1994.

\bibitem{Diosi}
L.~Di\'osi.
\newblock Models for universal reduction of macroscopic quantum fluctuations.
\newblock {\em Phys. Rev. A}, 40:1165--1174, Aug 1989.

\bibitem{Penrose}
Roger Penrose.
\newblock {On Gravity's role in Quantum State Reduction}.
\newblock {\em General Relativity and Gravitation}, 28(5):581--600, 1996.

\bibitem{Bassi}
Angelo Bassi, Andr{\'{e}} Gro{\ss}ardt, and Hendrik Ulbricht.
\newblock Gravitational decoherence.
\newblock {\em Classical and Quantum Gravity}, 34(19):193002, sep 2017.

\bibitem{Caves}
Carlton~M. Caves.
\newblock Quantum mechanics of measurements distributed in time. a
  path-integral formulation.
\newblock {\em Phys. Rev. D}, 33:1643--1665, Mar 1986.

\bibitem{Mensky}
M.~B. Mensky.
\newblock {\em Continuous Quantum Measurements and Path Integrals}.
\newblock Boca Raton: CRC Press, 1993.

\bibitem{Nelson}
Edward Nelson.
\newblock Derivation of the schr\"odinger equation from newtonian mechanics.
\newblock {\em Phys. Rev.}, 150:1079--1085, Oct 1966.

\bibitem{Guerra}
Francesco Guerra.
\newblock Structural aspects of stochastic mechanics and stochastic field
  theory.
\newblock {\em Physics Reports}, 77(3):263 -- 312, 1981.

\bibitem{Jacobson}
Ted Jacobson.
\newblock Thermodynamics of spacetime: The einstein equation of state.
\newblock {\em Phys. Rev. Lett.}, 75:1260--1263, Aug 1995.

\bibitem{Verlinde}
Erik Verlinde.
\newblock {On the origin of gravity and the laws of Newton}.
\newblock {\em Journal of High Energy Physics}, 2011(4):29, 2011.

\bibitem{Lloyd}
Seth {Lloyd}.
\newblock {The quantum geometric limit}.
\newblock {\em arXiv e-prints}, page arXiv:1206.6559, June 2012.

\end{thebibliography}
\end{document}